\documentstyle[aps,prl,preprint,floats,epsf]{revtex}

\begin{document}
\draft
\tighten

\preprint{\vbox{\hbox{\hfil CLNS 96/1452}
                        \hbox{\hfil CLEO 96-22}
}}

\title{Study of Gluon versus Quark Fragmentation in $\Upsilon \rightarrow
       gg\gamma$ and $e^+ e^- \rightarrow {q\bar q} \gamma$ Events at
	$\sqrt{s}$=10 GeV}


\maketitle

\begin{abstract}
Using data collected with the CLEO~II detector at the Cornell Electron Storage
Ring, 
we determine the ratio $R_{\rm chrg}$ 
for the mean charged multiplicity observed
in $\Upsilon({\rm 1S}) \rightarrow gg\gamma$ events, $<n_{\rm gluon}^{\pm}>$,
to the mean charged multiplicity observed in $e^+e^- \rightarrow q{\bar q}
\gamma$ events, $<n_{\rm quark}^{\pm}>$. We find
$R_{\rm chrg} \equiv \frac{<n_{gluon}^{\pm}>}{<n_{\rm quark}^{\pm}>}
= 1.04\pm 0.02 ({\rm stat})\pm 0.05 ({\rm syst})$ for jet-jet masses 
less than 
7 GeV.
\end{abstract}

\newpage

\begin{center}
M.~S.~Alam,$^{1}$ S.~B.~Athar,$^{1}$ Z.~Ling,$^{1}$
A.~H.~Mahmood,$^{1}$ H.~Severini,$^{1}$ S.~Timm,$^{1}$
F.~Wappler,$^{1}$
A.~Anastassov,$^{2}$ S.~Blinov,$^{2,}$%
\footnote{Permanent address: BINP, RU-630090 Novosibirsk, Russia.}
J.~E.~Duboscq,$^{2}$
% ...
\end{center}
 
\small
\begin{center}
$^{1}${State University of New York at Albany, Albany, New York 12222}\\
$^{2}${Ohio State University, Columbus, Ohio 43210}\\
% ...
\end{center}

\date{\today}
\tighten

\input psfig

{
 \renewcommand{\thefootnote}
 {\fnsymbol{footnote}}
 \setcounter{footnote}{0}
}

\newpage

\setcounter{footnote}{0}

\section{Introduction}
Understanding
hadronization, or the process by which elementary partons (gluons and quarks)
evolve into mesons and baryons, 
is complicated by its intrinsically
non-perturbative nature. Naively, one expects that because of the 
greater color charge of gluons compared to quarks, radiation of secondary
and tertiary gluons is more likely when hadronization is initiated by
a gluon rather than a quark. This results in a greater number of 
final state hadrons as well as a larger average opening angle between the
hadrons in the former case compared to the latter case. In the limit
$Q^2\to\infty$, the ratio of the number of hadrons produced in
gluon-initiated jets to the number of
hadrons produced in quark-initiated jets is expected, in
lowest order, to approach the
color degeneracy factor of 9/4\cite{r:theory76}.

Many experiments have searched for, and found,
multiplicity and jet shape differences between quark and
gluon fragmentation~\cite{r:Fuster}-\cite{r:OPAL6}.
At $Z^0$ energies, {\it q\=qg} events
are readily distinguished by their three-jet topology. Within such events,
quark and
gluon jets can be separated by a variety of techniques including vertex 
tagging.  Because 
gluons rarely fragment into heavy quarks, they will produce jets that 
form a vertex
at the $e^+e^-$ interaction
point. Quark jets, to the contrary, tend to form a detached
vertex when the jet contains
a long-lived bottom or charm quark.  
Unfortunately, the assignment of final state hadrons to the initial
state partons is rarely unambiguous and relies on Monte Carlo simulations
to determine the fraction of times that an observed hadron is correctly
traced to a primary parton.

The 10 GeV center of mass energy range offers a unique 
opportunity to
probe quark and gluon fragmentation effects, without relying on Monte
Carlo simulation to associate the final state hadrons with an initial state 
parton. The decay 
$\Upsilon(1S) \rightarrow gg\gamma$ allows one to compare the $gg$ system 
in a $gg\gamma$ event with the $q{\bar q}$ system in  
$e^+e^- \rightarrow q{\bar q}\gamma$ events. In both cases,
the system recoiling against the photon consists (to lowest order) of hadrons
that have evolved from either a two-gluon or a quark-antiquark system. 
The properties of the recoil systems can then be compared 
directly.\footnote{Although
there may be gluon radiation from the initial partons, we 
do not distinguish such radiation explicitly in this analysis.
Thus, the states that we are comparing are, strictly speaking,
$gg\gamma$ and $q{\bar q}\gamma$ to lowest-order only; 
additional gluon radiation, to which we are not experimentally sensitive,
may be present in many of the events in our sample.}
 
\section{Detector and Data Sample}

The CLEO~II detector\cite{r:CLEO-II} 
is a general purpose solenoidal magnet spectrometer and
calorimeter. 
The detector was
designed for efficient triggering and reconstruction of
two-photon, tau-pair, and hadronic events.
Measurements of charged particle momenta are made with
three nested coaxial drift chambers consisting of 6, 10, and 51 layers,
respectively.  These chambers fill the volume from $r$=3 cm to $r$=1 m, with
$r$ the radial coordinate relative to the beam ($z$) axis. 
This system is very efficient ($\epsilon\ge$98\%) 
for detecting tracks that have transverse momenta ($p_T$)
relative to the
beam axis greater than 200 MeV/c, and that are contained within the good
fiducial volume of the drift chamber ($|cos\theta|<$0.94, with $\theta$
defined as the polar angle relative to the beam axis). Below this 
threshold, the charged particle detection efficiency in the fiducial
volume decreases to 
approximately 90\% at $p_T\sim$100 MeV/c. For $p_T<$100 MeV/c, the efficiency
decreases roughly linearly to zero at a threshold of $p_T\approx$30 MeV/c.

Beyond the time of flight system is the electromagnetic calorimeter,
consisting of 7800 thallium doped CsI crystals.  The central ``barrel'' region
of the calorimeter covers about 75\% of the solid angle and has an energy
resolution of
\begin{equation}
\frac{ \sigma_{\rm E}}{E}(\%) = \frac{0.35}{E^{0.75}} + 1.9 - 0.1E;
                                \label{eq:resolution1}
\end{equation}
$E$ is the shower energy in GeV. This parameterization translates to an
energy resolution of about 4\% at 100 MeV and 1.2\% at 5 GeV. Two end-cap
regions of the crystal calorimeter extend solid angle coverage to about 95\%
of $4\pi$, although energy resolution is not as good as that of the
barrel region. 
The tracking system, time of flight counters, and calorimeter
are all contained 
within a 1.5 Tesla superconducting coil. 
Flux return and tracking
chambers used for muon detection are located immediately outside the coil and 
in the two end-cap regions.

We use 63 pb$^{-1}$ of
data collected at the $\Upsilon$(1S)
resonance ($\sqrt{s}$=9.46 GeV) as a source of 
$gg\gamma$ events and 198 pb$^{-1}$ at
$\sqrt{s}$=10.52 GeV (on the continuum just below the
$\Upsilon$(4S) resonance) as a source of
$q{\bar q\gamma}$ events. The $\gamma$ in our
$q{\bar q\gamma}$ sample results primarily from
initial state radiation (ISR)\cite{r:bkqed}.
We compare events for which the
invariant masses of the $gg$ and q\=q 
systems recoiling against the hard photon 
($M_{recoil}$, defined by $M_{recoil}$ =
$\sqrt{4E_{\rm beam}^{2}(1-E_{\gamma}/E_{\rm beam})}$)
are the same.

To suppress low-multiplicity QED events, 
we require that the 
thrust of the event (calculated using all the photon
candidates and good quality charged tracks)
be less than 
0.97, and that there be at least 3 good
charged tracks in the event.  Photon candidates are selected from showers 
with widths and patterns of energy deposition consistent with that of a 
photon, as opposed 
to neutral hadrons
(e.g. merged $\pi^0$'s, $K^0_{\rm L}$, neutrons, etc.). Owing to the 
excellent ability of the CLEO-II detector to distinguish
high-energy $\pi^0$'s from 
photons, approximately 75\% of the potential $\pi^0$ background is
removed at this stage by shower topology cuts. Additional $\pi^0$ suppression
is achieved with an explicit $\pi^0\to\gamma\gamma$ mass cut, 
to be discussed later.
To ensure that the events are well-contained within the CLEO detector,
we require 
$|\cos\theta_\gamma|<0.75$ ($\theta_\gamma$ is defined as before
as the 
   polar angle between the beam axis and the direct photon).

In addition to the
   $gg\gamma$ and $q\bar{q}\gamma$ samples, we search for 
$\Upsilon$(1S)$\to ggg$ and
   $e^+e^-\to q\bar{q}$ events containing a $\pi^0$ of energy comparable to the
   photon in the gamma-tagged sample.  These samples, referred to as
   $gg(\pi^0)$ and $q\bar{q}(\pi^0)$ are used to quantify the background
   levels from $\pi^0$ decays.

\section{Experimental Technique}

One of the most basic parameters used to characterize any event is the mean
charged track multiplicity, $<{\it N_{chrg}}>$.  We plot  $<{\it N_{chrg}}>$ 
as a
 function of the mass recoiling against the direct photon,
covering the 
recoil mass range from 4 to 7 $\rm{GeV}$.  This mass
interval corresponds to 0.45 $<  E_{\gamma}/E_{\rm beam} < 0.82$ for the 
$gg\gamma$ sample.
We require the recoil mass to be greater than 4 $\rm{GeV}$ to ensure 
that we 
are significantly
above the $c{\bar c}\gamma$ threshold; we expect that our 
$q{\overline q}$ jet sample therefore includes {\it u,d,s,c} 
quarks approximately
uniformly, apart from small phase space
effects.\footnote{We note that the q\=q mixture in
our q\=q$\gamma$ sample may be different for cases in which the photon is
emitted in the final state as opposed to the initial state due to the
different diagrams responsible for these processes. We expect our sample
to be dominated by ISR events, as is evidenced by the photon polar
angle distribution.}

To determine the characteristics of 
$gg\gamma$ events, we must subtract the background from non-resonant
$q{\bar q\gamma}$ and 
$e^+e^-\to\tau\tau\gamma$ events produced in $e^+e^-$ annihilations at
$\sqrt{s}=M_{\Upsilon{\rm 1S}}$. 
This can be done by direct scaling of the event sample collected at
the continuum of the 
$\Upsilon$(4S)
resonance.
In addition to $q{\bar q\gamma}$ events,
this latter sample of events also
includes $\tau\tau\gamma$ events, as well as any other 
continuum backgrounds which
may be present in our resonant sample.
By subtracting this ``raw'' sample from our $\Upsilon$(1S) sample, 
we therefore account for all of the non-resonant 
backgrounds to $gg\gamma$ at the 
$\Upsilon$(1S) energy.  
We find that before subtraction,
this background to $gg\gamma$ comprises about 
$10\%$ of the gamma-tagged sample taken at the
$\Upsilon$(1S) energy.

Similarly, in order to isolate 
$q{\bar q\gamma}$ events 
at 10.52 $\rm{GeV}$, we must quantify $\tau\tau\gamma$ contamination.  
This is
done using a Monte Carlo simulation of tau pair events.  We find that  
$\tau\tau\gamma$ events comprise about $10\%$ of the $q{\bar q\gamma}$
data sample passing the other event selection cuts specified above.  Beam 
gas and two photon backgrounds were investigated and found 
to be negligibly small.



To determine the level 
of $gg(\pi^0)$ contamination to $gg\gamma$ and $q{\bar q}(\pi^0)$
contamination to $q{\bar q}\gamma$, we 
first determine how often a single high energy photon 
can be matched with other photons in the same event to form a $\pi^0$.  Such
photons are explicitly vetoed as likely $\pi^0$ daughters. The 
$\pi^0$ contamination, of course, results from cases for which
we find only
one of the two $\pi^0$ daughter photons in the detector\cite{r:hancock}.
From Monte Carlo simulations,
we determine the probability 
that the second $\pi^0$ daughter photon will be found. 
The likelihood of detecting 
the second photon varies from $50\%$ 
at $M_{recoil}= 6.5$ $\rm{GeV}$ to $77\%$ 
at $M_{recoil} = 4$ 
$\rm{GeV}$. 
Knowing the fraction of times that the second daughter photon goes
undetected, and the total number of $\pi^0$'s that we reconstruct in our
sample, we thereby 
determine the fraction of our direct photon candidates 
which are actually
$\pi^0$ daughters, but are not vetoed as such. This is shown in
(Figure~\ref{fig:pi0contam}) as a function of recoil mass.

Our backgrounds are summarized below:

\begin{center}
\begin{tabular}{c|c|c}
\hline
Sample & $\sqrt{s}$ & Background ($\%$) \\
\hline
\hline
$gg\gamma$ & 9.46 GeV & Sum of all non-$\Upsilon$(1S) events ($10\%$)  \\
$gg\gamma$ & & $\Upsilon$(1S)$\to gg(\pi^0)$ 
contamination to $\Upsilon$(1S)$\to gg\gamma$ ($\sim 5\%$)  \\
\hline
$q{\bar q\gamma}$ & 10.52 GeV & $\tau\tau\gamma$ ($10\%$) \\
$q{\bar q\gamma}$ & & $e^+e^-\to q{\bar q}(\pi^0)$ contamination to
$q{\bar q}\gamma$ ($\sim 8\%$) \\
$q{\bar q\gamma}$ & & $\gamma\gamma$ ($<1\%$) \\
$q{\bar q\gamma}$ & & beam-gas ($<1\%$) \\
\hline
\end{tabular}
\end{center}

\begin{figure}
\centering
\centerline{\hbox{\psfig
{file=3031296-012.ps,width=5.75truein,bbllx=75bp,bblly=176bp,bburx=537bp,bbury=696bp}}}
\vskip 2.0cm
\caption[]
{$\pi^0$ contamination, as a function of recoil mass, in $gg\gamma$ 
events
 (squares), and in $q{\bar q\gamma}$ events (circles); the y axis gives the fraction of our 
apparent
 $gg\gamma$ (or $q{\bar q}\gamma$)
event sample which are actually $gg(\pi^0)$
 (or $q{\bar q}(\pi^0)$) events.}
\label{fig:pi0contam}
\end{figure}

\begin{figure}[p]
\centering
\centerline{\hbox{\psfig
{file=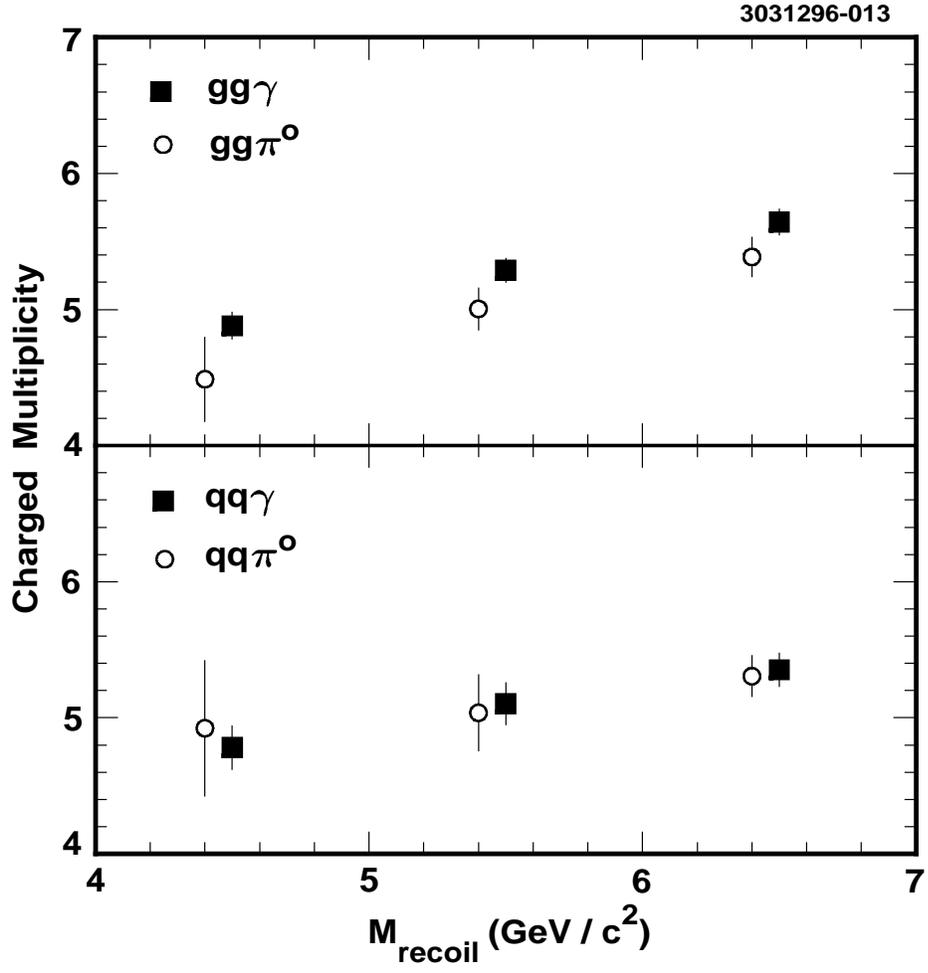,width=5.75truein,bbllx=59bp,bblly=176bp,bburx=521bp,bbury=626bp}}}
\vskip 2cm
\caption[]
{A comparison of the charged multiplicity distributions for $gg\gamma$ and 
 $gg(\pi^0)$ (top) and $q{\bar q\gamma}$ and $q{\bar q}(\pi^0)$ (bottom).}
\label{fig:gggam/ggpi0nchrgcompare} 
\end{figure}

To determine how our measured charged multiplicity values are biased by the
$\pi^0$ background, we make
a comparison plot of the charged multiplicity distribution for $gg\gamma$ 
vs. $gg(\pi^0)$ events, and a similar plot for $q{\bar q\gamma}$ vs. 
$q{\bar q}(\pi^0)$ events.  From Figure~\ref{fig:gggam/ggpi0nchrgcompare}, we 
see that the charged multiplicity distributions for 
the gamma-tagged and the $\pi^0$-tagged samples
are similar.\footnote{It is of some interest to note the similarity in
multiplicity between the $gg\gamma$ and the $gg(\pi^0)$ samples. 
Although the first process contains two primary gluons, whereas the
second contains three primary gluons,
this plot
qualitatively suggests that at these energy scales
the available energy for fragmentation
is the main factor in determining the final state charged multiplicity.} 
Quantitatively, the effect of $\pi^0$ backgrounds is to reduce the measured
value of $R_{chrg}$ by about 1\% relative to the true value. We
statistically correct for this effect in subsequent plots, and
include the 
uncertainty due to $\pi^0$ backgrounds in our overall systematic error.

\begin{figure}[p]
\centering
\centerline{\hbox{\psfig
{file=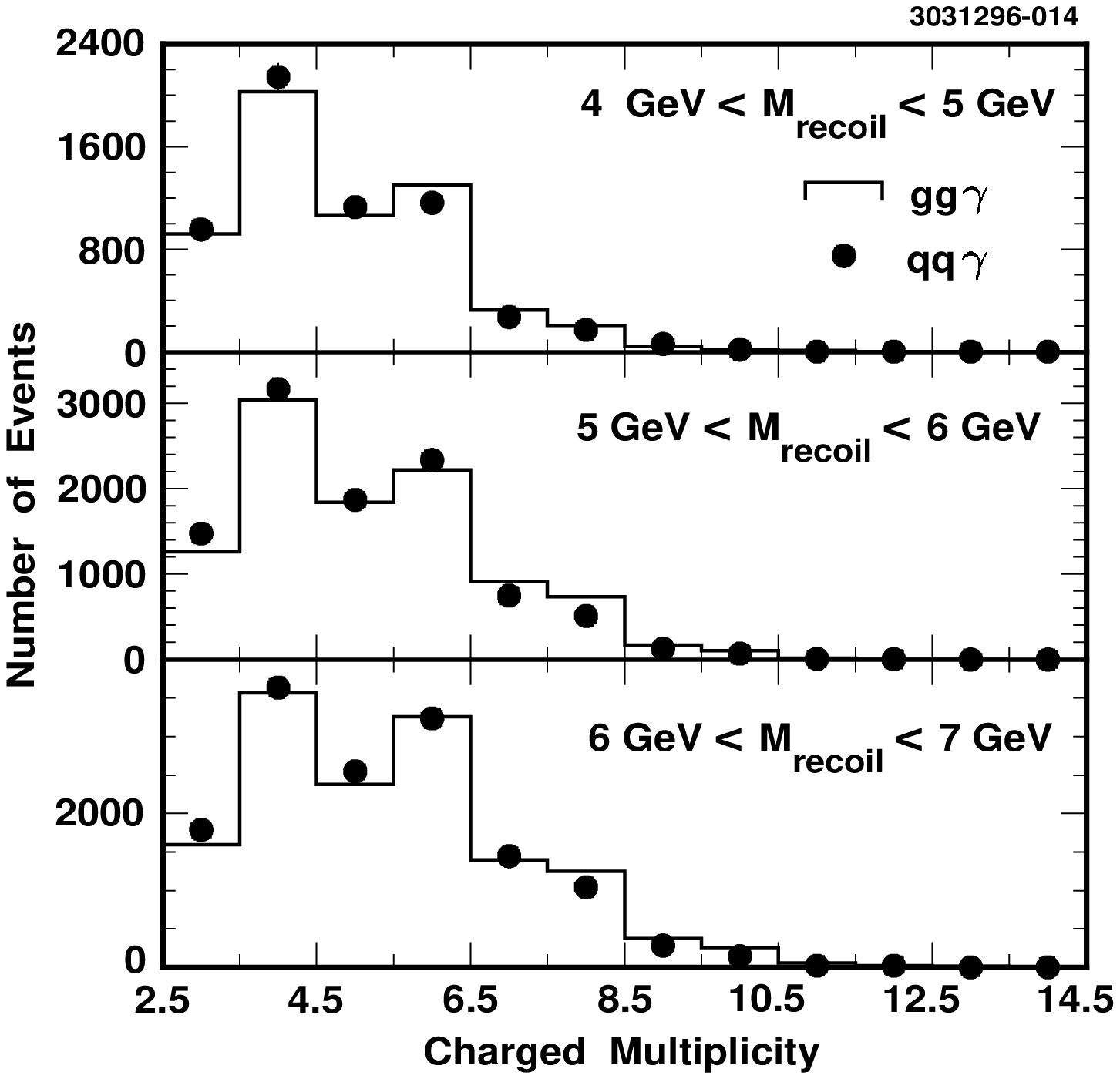,width=5.75truein,bbllx=88bp,bblly=190bp,bburx=531bp,bbury=662bp}}}
\vskip 2cm
\caption[]{Observed multiplicity distributions for background subtracted
 $gg\gamma$ and $q{\bar q\gamma}$ data for different recoil mass ranges.}
\label{fig:mult dist.} 
\end{figure}

\begin{figure}[p]
\centering
\centerline{\hbox{\psfig
{file=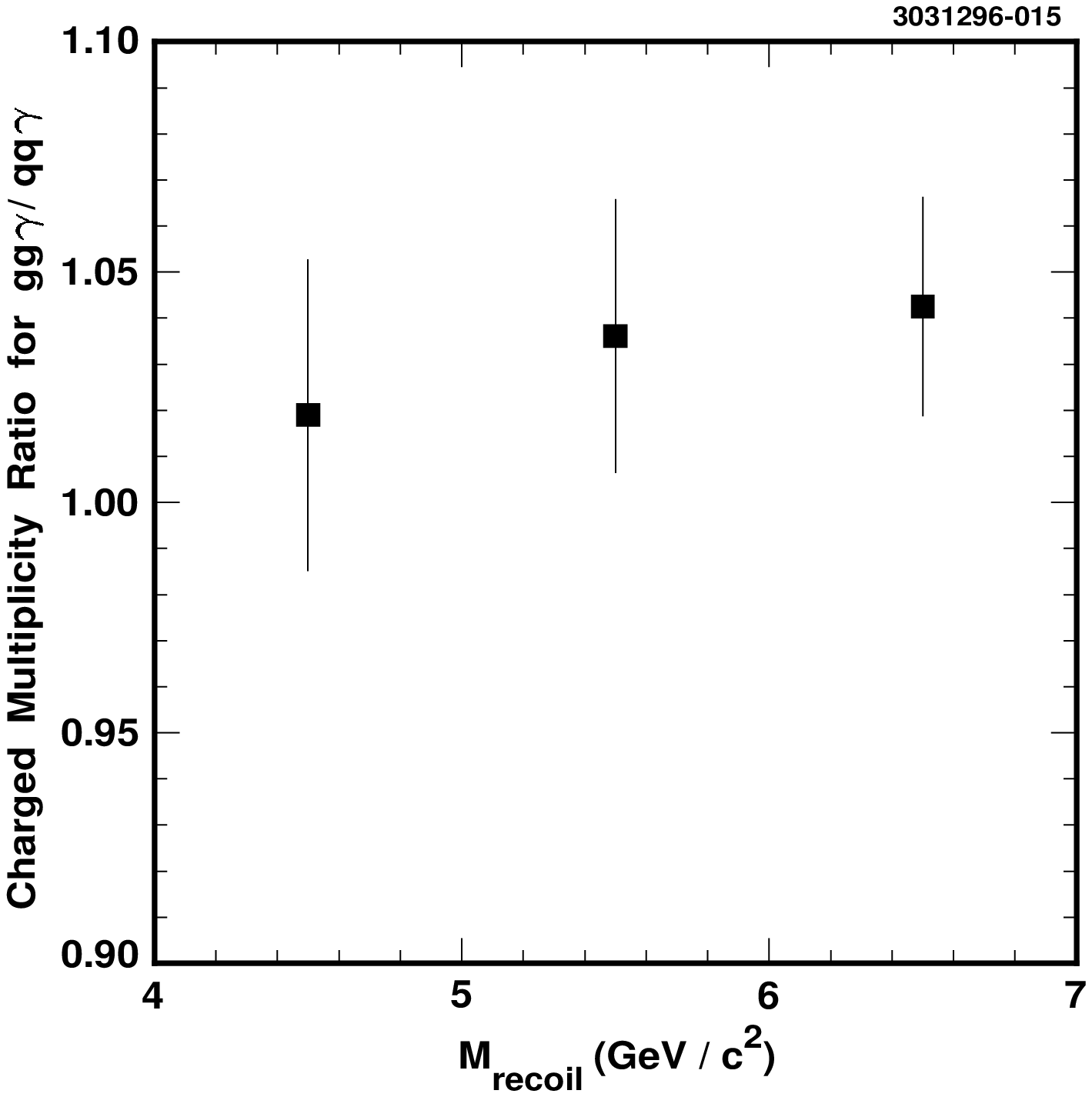,width=5.75truein,bbllx=88bp,bblly=190bp,bburx=531bp,bbury=662bp}}}
\vskip 2cm
\caption[]{
 Charged multiplicity ratio for background subtracted 
$gg\gamma$/$q{\bar q\gamma}$ data.}
\label{fig:subtracted plots} 
\end{figure}

In Figure~\ref{fig:mult dist.}, we show the multiplicity distributions in 
bins of recoil mass
for our background-corrected
$gg\gamma$ and $q{\bar q\gamma}$ samples.  From the
distributions in this Figure, we
extract the mean multiplicities as a function of {\it gg}
 or $q{\bar q}$ mass, for pure $gg\gamma$ and $q{\bar q\gamma}$ 
samples.  As shown in 
Figure~\ref{fig:subtracted plots}, the ratio of  $<{\it N_{chrg}}>$ resulting 
from
gluon fragmentation to  $<{\it N_{chrg}}>$ from quark fragmentation is
$R_{chrg} = 1.04\pm0.02$, after all the aforementioned background corrections.

\section{Systematic errors}


In order to investigate the sensitivity of the result to the 
${\it N_{chrg}}\ge 3$ requirement and to suppress 
$\tau\tau\gamma$ backgrounds, we compare the ratios obtained
from ${\it N_{chrg}}\ge 3$ with ratios obtained using the 
tightened 
multiplicity interval 5$\le {\it N_{chrg}}\le$11.  
This comparison also 
partially addresses possible biases in our measured ratio due to 
acceptance effects as well as the effect of the minimum charged multiplicity
cut. If the shapes of the true charged multiplicity distributions were much
different for quarks vs. gluons at low multiplicities (for which our
acceptance is worse than at high multiplicity, due to the effect of
our hadronic event selection cuts as well as trigger effects), then
our derived average will be artificially `pulled' by our $N_{chrg}\ge$3
requirement. A
complete Monte Carlo simulation study of possible differences between
the detector level multiplicities and 
the true multiplicities, including unfolding effects, indicates 
negligible bias. The value of the ratio using $N_{chrg}>$5 is found to be
statistically consistent with that for $N_{chrg}>$3.
The effect of imposition of the minimum $N_{chrg}$ cut has 
also been checked by repeating the entire analysis, but using the neutral
multiplicity $N_{neut}$ rather than the charged multiplicity as the comparison
variable. For this cross-check, we retain the cut $N_{chrg}\ge 3$,
but make no cut on the minimum neutral multiplicity.
Before any consideration of backgrounds, this cross-check yields 
$R_{neut}=1.06\pm0.03$ (statistical error only).
This value is in good agreement with our measured value of $R_{chrg}$.


Although we compare systems of equivalent mass, the energies of the systems 
recoiling against the photon are higher for $q\bar q$ events than for $gg$ 
events at an
equivalent invariant mass.  On average, the $q{\bar q\gamma}$ system will 
therefore be 
more collimated.  We investigate possible energy dependent 
effects with Monte Carlo simulation, by looking at systems of equal 
recoil mass, but generated at
different energies ($\sqrt{s}$=10.52 GeV vs.
$\sqrt{s}$=9.46 GeV).  
We find a systematic shift of 0.006 in the ratios.  
We can also evaluate energy effects by direct inspection of
Figure \ref{fig:subtracted plots}.
Although the $R_{chrg}$ values measured for the 
lowest recoil mass are statistically consistent with the
$R_{chrg}$ values measured for the highest recoil mass
considered, our results are also consistent with 
$R_{chrg}$ increasing slightly with recoil mass.
We conservatively include a systematic error
of 2\% to account for any possible systematic variation.

We also compare the $gg\gamma$/$q{\bar q\gamma}$ multiplicity  
ratios from data taken during three 
distinct running periods, covering different running conditions and
triggers.
The separate data sets give ratios which are all statistically
consistent with each other. 
Nevertheless, we conservatively include the
maximum variation of $2\%$ observed
between the 
three data sets
in the overall systematic error.

Finally,
we consider the dependence of the mean multiplicity on the dip 
angle, $|\cos\theta_\gamma|$, of the direct photon.
We expect that initial state radiation
$q{\bar q\gamma}$ events tend to produce photons that are more forward peaked
(have larger values of $|\cos\theta_\gamma|$) than for $gg\gamma$.  For   
$q{\bar q\gamma}$ events, we therefore expect that the 
particles emerging
on the other side of the event, opposite the high energy $\gamma$, are more 
likely to be lost down the beam pipe.  
We have therefore explicitly considered any possible dependence of
$R_{chrg}$ on $|cos\theta_\gamma|$ by plotting $R_{chrg}$ in different
%
%
bins of 
$|\cos\theta_\gamma|$. 
At our level of statistical precision, we do not
observe any such systematic effect and conclude
that the measured $R_{chrg}$ value is not 
measurably biased by such
acceptance effects.\footnote{There are potentially two competing effects here
- although the photon is more forward peaked for $q{\bar q}\gamma$ events, it
is also the case that quark jets should be more narrowly collimated than
$gg\gamma$ events. This may tend to mitigate any effects due to the difference
in photon
angular distributions, since charged hadrons produced in $q{\bar q}\gamma$
events should be more contained within
the fiducial region of our
drift chamber.} We explicitly quantify this by measuring $R_{\rm chrg}$
in two regions of $cos\theta_\gamma$: $|cos\theta_\gamma|<$0.35, and 
$0.35<|cos\theta_\gamma|<$0.70. The variation between the two ratios is
found to be $0.02\pm 0.03$. We conservatively assign the error on
this value
as a measure of the maximum possible
systematic error due to such polar angle acceptance effects.


In summary, the systematic errors are:

\begin{center}
\begin{tabular}{c|c}
\hline
Systematic Error & Value \\ \hline
\hline
$\pi^0$ background & 0.003 \\
\hline 
Recoil mass-dependent effects & 0.006 \\
\hline
Truncated (5$\le {\it N_{chrg}}\le$11) vs. non-truncated mean & 0.02 \\
\hline
Variation between data sets & 0.02 \\
\hline
$R_{chrg}$ dependence on $M_{\rm recoil}$ & 0.02 \\
\hline
$|\cos\theta|$ cut dependence & 0.03 \\
\hline
\hline
Total & 0.05 \\
\hline
\end{tabular}
\end{center}

\section{SUMMARY}
The ratio of  $<$${\it N_{chrg}}$$>$ for 
gluons to  
$<$${\it N_{chrg}}$$>$ for quarks measured here is smaller than those 
found by the OPAL, 
ALEPH, SLD and DELPHI experiments, at $\sqrt{s}\sim M_{Z^0}$.
  The ratios compare as follows:

\begin{center}
\begin{tabular}{c|c|c}
\hline
Collaboration & $<$$N$$>_g / <$$N$$>_q$ & Kinematic Regime \\
\hline
CLEO 96 & $1.04\pm0.05$ & $<E_{jet}>~ <~$7 $\rm{GeV}$ \\
\hline
DELPHI~\cite{r:DELPHI} & $1.24\pm0.015$ & $<E_{jet}>$ = 10-40 $\rm{GeV}$ \\
\hline
DELPHI~\cite{r:DELPHI} & $1.06\pm0.18$ & $<E_{jet}>$ = 10 $\rm{GeV}$ \\
\hline
SLD~\cite{r:SLD} & $1.36\pm0.24$ & $<E_{jet}>$ = 24  $\rm{GeV}$ \\ 
\hline
OPAL 93~\cite{r:OPAL2} &  $1.27\pm0.04$ & $<E_{jet}>$ = 24  $\rm{GeV}$ \\
\hline
ALEPH~\cite{r:ALEPH2} & $1.19\pm0.04$ & $<E_{jet}>$ = 24  $\rm{GeV}$ \\
\hline
OPAL 96~\cite{r:OPAL5}$^*$ & $1.58\pm0.03$ & $<E_{jet}>$ = 39  $\rm{GeV}$ \\
\hline
\end{tabular}
\end{center}

\begin{center}
$^*$This OPAL result is for $<N>_{g}-1 / <N>_{q}-1$, considering only {\it uds}
 quarks
\end{center}

Fuster and Mart\'{i}~\cite{r:Fuster}, DELPHI~\cite{r:DELPHI} and 
Fodor~\cite{r:Fodor} have recently 
considered how the ratio of charged 
multiplicity for gluon to quark jets depends on the hadronic 
center-of-mass energy. Phenomenologically, as the energy scale increases,
the likelihood of successive partons radiating also increases.
Having a greater color
degeneracy than quarks, the multiplicity of gluon-initiated jets increases
faster than the multiplicity of quark-initiated jets.  This causes 
$R_{chrg}$ to increase slowly with energy.  Our result is
smaller than the LEP results for $R_{chrg}$, showing
the expected energy dependence\cite{r:explanation}.
Our result is also consistent with
the 
measurement by DELPHI at 
$\sqrt{s}\sim$10 GeV, albeit with substantially better precision.
We note that the $q{\bar q}$ 
mixture in this experiment is likely to be different than the quark
mixture for the high energy 
experiments mentioned above.
Recent analyses by
OPAL~\cite{r:OPAL6} and ALEPH~\cite{r:ALEPH2}, in fact, have explicitly 
measured the difference in ratio values
when comparing gluon jets to b-quark
jets versus gluon jets to uds-quark jets.  
\section{Acknowledgments}

We gratefully acknowledge the effort of the CESR staff in providing us with 
excellent luminosity and running conditions.  J.P.A., J.R.P.,
and I.P.J.S. thank the NYI program of the NSF, G.E. thanks the Heisenberg
Foundation, K.K.G., M.S., H.N.N., T.S., and H.Y. thank the OJI program of the 
DOE, J.R.P., K.H., and M.S. thank the A.P. Sloan Foundation, L.J.W. thanks
the Hughes Foundation, and A.W. and R.W. thank the Alexander von Humboldt 
Stiftung for support.  This work was supported by the U.S. National Science 
Foundation, the U.S. Department of Energy, and the Natural Sciences and 
Engineering Research Council of Canada. 

\newpage



\begin{thebibliography}{99}
\bibitem{r:explanation}{Fuster and Mart\'{i}, for example, predict 
$R_{chrg}=0.988\pm 0.122$ for $E_{jet}$=10 GeV. They do not extrapolate as 
low as the energies probed in this measurement, however.}~\cite{r:private}
\end{thebibliography}
\end{document}